\documentclass[conference]{IEEEtran}

\IEEEoverridecommandlockouts

\ifCLASSINFOpdf
\else
\fi

\usepackage{amsmath}

\usepackage{algorithm,algorithmic}

\ifCLASSOPTIONcompsoc
  \usepackage[caption=false,font=normalsize,labelfont=sf,textfont=sf]{subfig}
\else
  \usepackage[caption=false,font=footnotesize]{subfig}
\fi

\usepackage{graphicx}
\usepackage{tabularx} 
\usepackage{color}
\usepackage{float} 
\usepackage{paralist} 
\usepackage[T1]{fontenc}
\usepackage{currvita}
\newcolumntype{b}{X}
\newcolumntype{s}{>{\hsize=.5\hsize}X}
\usepackage{amsfonts}

\usepackage{epstopdf}
\begin{document}

\title{On the Impact of Wireless Jamming on the\\ Distributed Secondary Microgrid Control}

\author{Pietro Danzi\IEEEauthorrefmark{1}, \v{C}edomir Stefanovi\'c\IEEEauthorrefmark{1}, Lexuan Meng\IEEEauthorrefmark{2}, Josep M. Guerrero\IEEEauthorrefmark{2}, Petar Popovski\IEEEauthorrefmark{1}\\
\IEEEauthorrefmark{1}Department of Electronic Systems, Aalborg University, Denmark, 
Email:\{pid,cs,petarp\}@es.aau.dk \\
\IEEEauthorrefmark{2}Department of Energy Technology, Aalborg University, Email:\{lme,joz\}@et.aau.dk }

\maketitle

\begin{abstract}
The secondary control in direct current microgrids (MGs) is used to restore the voltage deviations caused by the primary droop control, where the latter is implemented locally in each distributed generator and reacts to load variations.
Numerous recent works propose to implement the secondary control in a distributed fashion, relying on a communication system to achieve consensus among MG units.
This paper shows that,  if the system is not designed to cope with adversary communication impairments, then a malicious attacker can apply a simple jamming of a few units of the MG and thus compromise the secondary MG control.
Compared to other denial-of-service attacks that are oriented against the tertiary control, such as economic dispatch, the attack on the secondary control presented here can be more severe, 
as it disrupts the basic functionality of the MG.
\end{abstract}


\IEEEpeerreviewmaketitle

\section{Introduction}
\label{intro}
Microgrid (MG) control is typically organized according to a hierarchical architecture \cite{guerr1} to differentiate among control objectives, as each of them has different requests in terms of time scale and quantity of information needed.
The \emph{primary control} adjusts the electrical parameters based only on local measurements. 
A common approach for the primary control is to use the droop method, which is simple to implement, but at the same time causes the grid voltage deviation.
The \emph{secondary control} is used to compensate these deviations and can be implemented in a centralized or distributed fashion, where in both cases a support of a communication system is required.
Finally, the optimization functionalities, such as, e.g., economic dispatch, are grouped in the \emph{tertiary control}.

There is a growing interest in the MG research community to use wireless communications for achieving coordination among Distributed Generators (DGs), due to their plug-and-play capability and greater flexibility compared to a cabled system.
Another favorable point is the broadcast nature of the communication, that permits the implementation of distributed coordination algorithms \cite{porco}-\cite{nasirian}.
The benefits of a distributed approach are the absence of a single point of failure and lower bandwidth requirements.
The suitable communication technologies for MG control are preferred to be low-cost and easy to implement, such as, e.g., Wi-Fi or ZigBee \cite{cady}.

The use of a wireless technology exposes the system to multiple security attacks, notably the jamming attacks. 
In this work we study the performance of distributed secondary MG control supported by WiFi, i.e., 802.11 networking, in case of jamming attack.
Specifically, we show that a simple jamming strategy, which exploits only the periodicity of the communication among DGs and attacks just a subset of DGs, leads to a failure of the voltage restoration.
We also present possible countermeasures to this attack.

The organization of the rest of the text is as follows.
Section~\ref{sec:background} contains a classification of denial-of-service (DoS) attacks and the related work in context of MG control.
In Section~\ref{sec:model} we present the MG model, consisting of an electrical part, distributed control and communication protocol, as well as models of the interference and the jammer.
In Section~\ref{sec:analysis} we analyze the overall system and state the control objective.
The impact of the jamming and the potential countermeasures are presented in Section~\ref{sec:attack}.
Section~\ref{sec:conclusion} concludes the paper.

\section{Background and Related Work}
\label{sec:background}

Various DoS attacks can be conducted against a 802.11 network by a jammer, which is defined as \emph{an entity who is purposefully trying to interfere with the physical transmission and reception of wireless communications} \cite{xu}.
The DoS attacks by jammers are classified as basic and intelligent \cite{pelechrinis}.
Basic jammers transmit a jamming signal either constantly, randomly or as \emph{reactive} jammers, which are activated only during the transmissions that are intended to be jammed.
In this way, the jammer reduces energy consumption and the probability of being detected.
This attack results in back-off freezing of all the stations that are in the jammer transmission range.
On the other hand, an intelligent jammer can try to corrupt the control messages, such as acknowledgment packets.
If the jammer is capable of packet forging, it can perform more sophisticated attacks exploiting protocol weaknesses, such as the deauthentication attack, the virtual carrier sense attack, or cryptographic-caused DoS, cf. \cite{eian}.

The importance of the resilience to DoS attacks has been already underlined in the smart grid literature \cite{wang}.
Specifically, the attacks can be conducted against the \emph{confidentiality}, \emph{integrity} and \emph{availability} of the communication channel: we are interested in the availability of the communication for the distributed secondary MG control.
Some related works show countermeasures to attacks against the economic dispatch or other tertiary control objectives, e.g., a regret matching based anti-jamming algorithm is considered in \cite{chai}.
Regarding the secondary control, it is noted in \cite{friedberg} that the most critical phase is the change of operation mode, but the considered system has a centralized control.
In this work, we show that an attack to the distributed secondary control leads to more severe consequences and can undermine the grid stability.

\section{System Model}
\label{sec:model}

\begin{figure}[!tb]
\centering

\subfloat[Microgrid with 6 DGs.]{\includegraphics[width=\columnwidth]{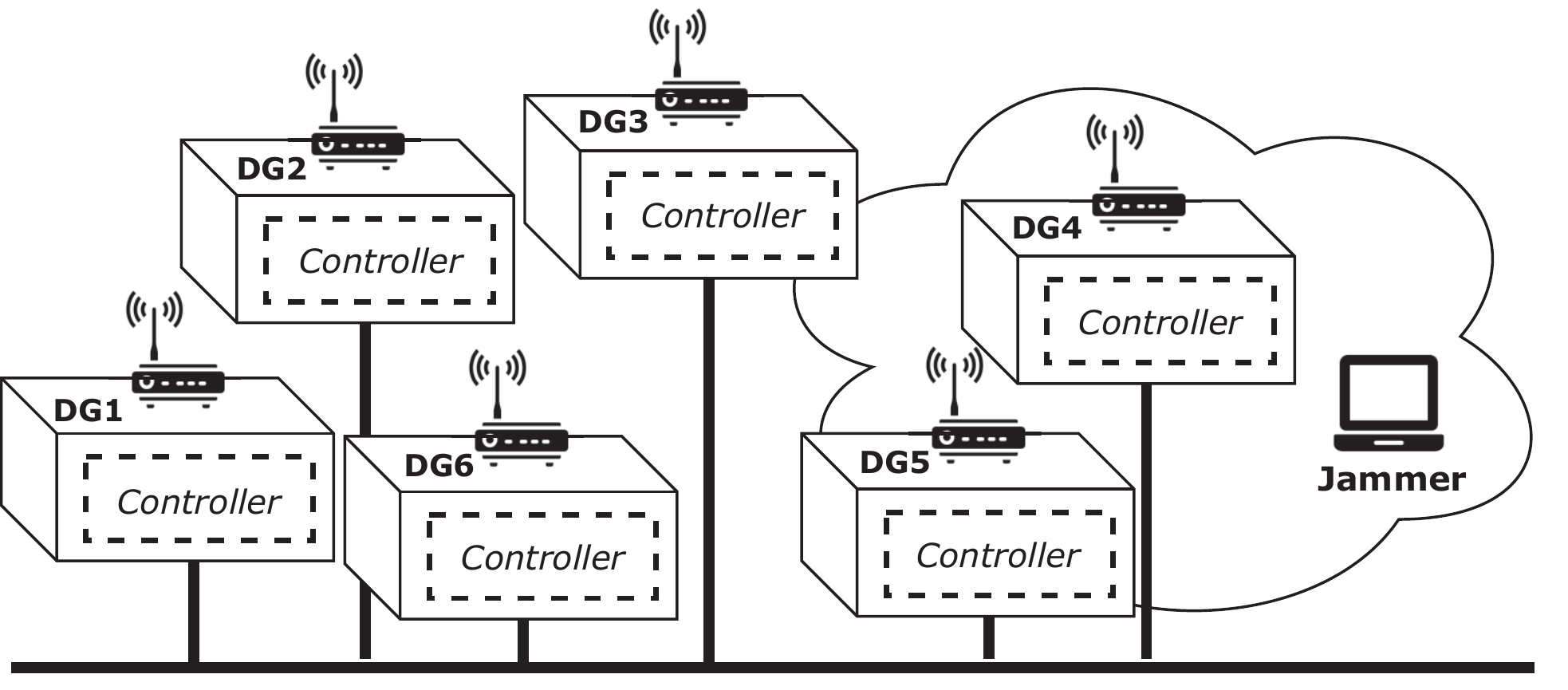}}
\hfil
\subfloat[Schematic of the control loop of a DG.]{\includegraphics[width=\columnwidth]{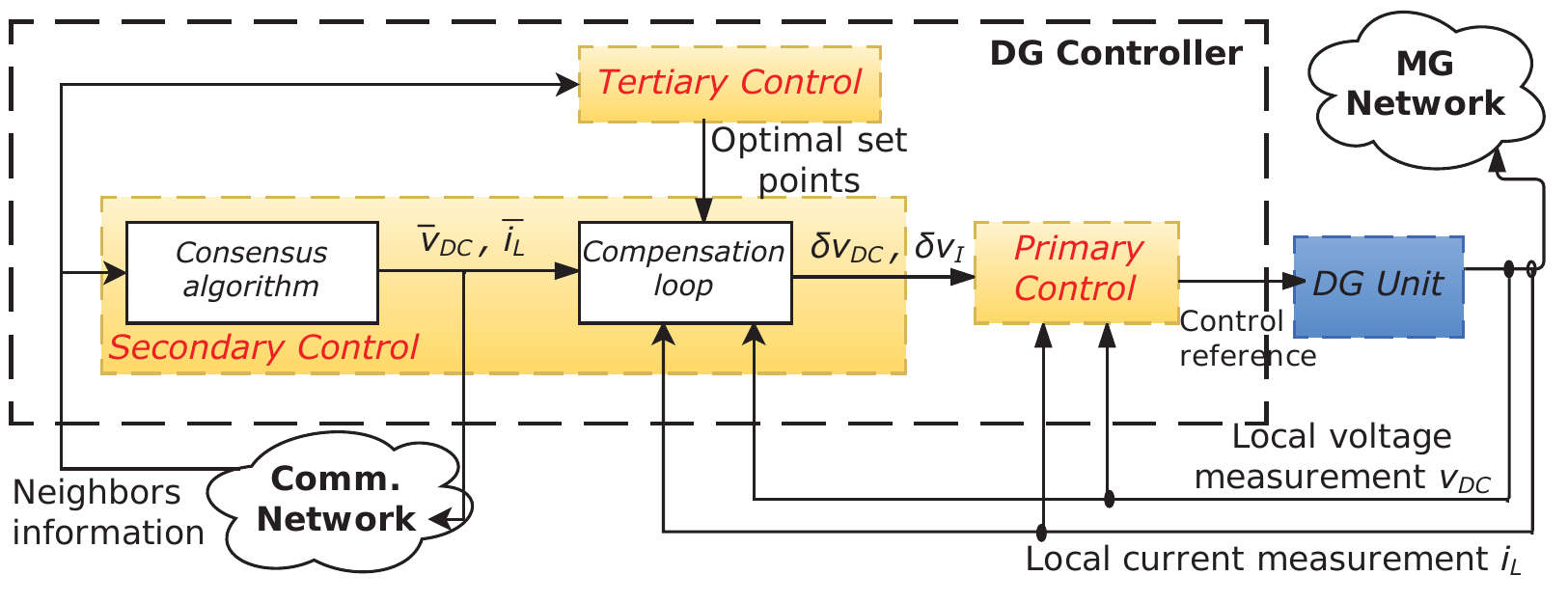}}
    \caption{System model.}
    \label{fig:control_loop}
    \vspace{-6pt}
\end{figure}

We consider a residential DC microgrid operating in an islanded mode (i.e., disconnected from the main grid), inspired by a reference model from \cite{meng2} and depicted in Fig.~\ref{fig:control_loop}(a).
There are $N=6$ DG units interconnected via electrical distribution lines and regulated by a hierarchical controller, Fig.~\ref{fig:control_loop}(b).

\subsection{Control Model}
Although the primary control roughly distributes the power among DGs, its accuracy largely depends on electrical parameters and the voltage deviations appear.
The secondary control is used to eliminate these deviations by generating two compensation terms $\delta v_I$ and $\delta v_{DC}$ that are fed back to the primary controller:
\begin{align}
\delta v_I & = \left( \frac{K_{isc}}{s} + K_{psc} \right) \cdot ( \bar{i}_L - i_L ), \\
\delta v_{DC} & = \left( \frac{K_{psv}}{s} + K_{isv} \right) \cdot ( V_r - \bar{v}_{DC} ),
\end{align}
where $\bar{i}_L$ and $\bar{v}_{DC}$ are the estimated average values of the output currents and voltage, $V_r$ is the voltage reference value, $K_{isc}$, $K_{isv}$ and $K_{psc}$, $K_{psv}$ are the integral and proportional terms of the PI controllers, respectively, and $s$ is the Laplace operator \cite{meng2}.

\begin{figure}[!tb]
\minipage{\columnwidth}
\centering
  \includegraphics[width=0.8\columnwidth]{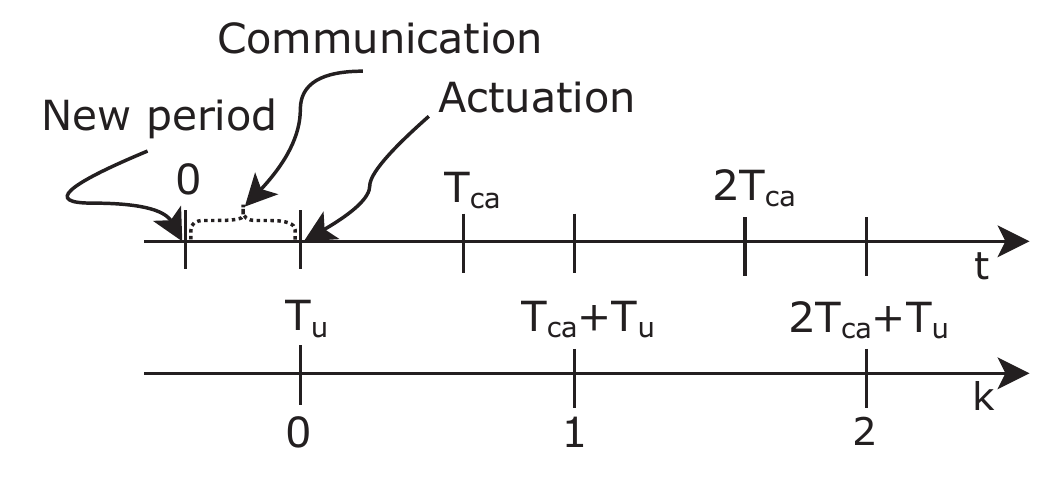}
\endminipage\hfill
    \caption{Time scale and consensus sampling period.}
    \label{fig:time_sample}
\end{figure}

In order to obtain the estimated average values, a consensus algorithm can be implemented in DG controllers \cite{meng}, which are equipped with 802.11 radio interfaces.
To a networked DG we refer as an \emph{agent}, following the nomenclature used in the literature.
The correction terms are computed locally every $T_{ca}$ seconds via the consensus algorithm, which operates on the information sent by agents that are in the communication range.
Specifically, in each interval of $T_{ca}$, there is a subinterval of length $T_{u}$, $T_u \leq T_{ca}$, dedicated to the exchanges of the status updates among the units (i.e., communication), after which the collected information are given as input to the consensus block, see Fig.~\ref{fig:time_sample}.
Thus, the consensus is a periodic process with steps $k T_{ca} + T_u$, $ k \geq 0 $.
Ideally, if all communication updates are received within $T_u$, the output of the consensus is
\begin{equation}\label{ca}
\mathbf{x}_i(k+1) = W_{ii} \, \mathbf{x}_i(k) + \sum_{j \in \mathcal{N}_i} W_{ij} (\mathbf{x}_j(k)-\mathbf{x}_i(k)),
\end{equation}
where $\mathbf{x}_i = [\bar{i}_{L,i} , \bar{v}_{DC,i}]$ is the state of the $i$-th agent, $W_{ij}$ are the weight coefficients, and $\mathcal{N}_i$ denotes the set of the neighbors of $i$-th agent.
In this work we assume that $W_{ii} = 1$, $\forall i$, and $W_{ij} = \epsilon$, $ i \neq j$, i.e., $W_{ij}$ are constant, as in \cite{xiao}.

\subsection{Communication Model}

The communication objective is to periodically deliver the agent state information to neighbors in a reliable way.
The communication graph is assumed to be connected and, thus, allows for the convergence of the consensus.
We assume that there are two impairments to communication reliability: (i) unintentional interference from, e.g., a domestic access point that provides services to the inhabitants, and (ii) intentional interference by the attacker (jammer) that is located in proximity of the agents, with a goal of compromising the MG communications.
The agents are communicating with a low and fixed modulation scheme, and the length of the packet payload exchanged by the agents is also constant.
Therefore, the on-air time of a packet is constant and assumed to be equal to $T_p$.
The agents are broadcasting packets using the medium access control regulated by the 802.11 distributed coordination function (DCF).

\subsubsection{IEEE 802.11 DCF}

The DCF mechanism in the IEEE 802.11 standard is a carrier sense multiple access with collision avoidance (CSMA/CA) and exponential back-off.
The carrier sensing is performed by the clear channel assessment (CCA) module that reports if the channel is busy or idle.
When a station has a packet in the transmission queue, it starts sensing the channel status.
If the channel is sensed idle for a distributed-inter-frame-space (DIFS) interval, the packet is sent.
Otherwise, it starts the back-off procedure.
A back-off value is chosen randomly from $[0,W-1]$, where $W$ is the maximum back-off value.
The station is allowed to transmit only when the back-off counter reaches zero and the channel is sensed idle.
The channel state is updated every carrier sense (CS) time-slot of $\sigma$ seconds by the CCA.
The back-off value is decremented for each idle time slot.
If the channel is busy, the back-off is frozen until the channel is sensed idle for a DIFS, after which decrementing continues.
After a successful transmission, the 802.11 DCF starts a new back-off procedure, even if there are no enqueued packet; this mechanism is referred to as post back-off. 
Finally, in case of broadcast packets there is no acknowledgement from the receiver and the value of the back-off window $W$ is fixed.

The delay from the instant that a packet is sent to the interface of a transmitting agent to the instant that its payload is disposable for the secondary control of a receiving agent is:
\begin{equation}
T_{MAC} = T_{queue} + T_{DCF} + T_p + T_{rx},
\end{equation}
where $T_{queue}$ is the time spent in the transmission queue, $T_{DCF}$ is the time consumed by the contention mechanism, $T_p$ is the on-air time, and $T_{rx}$ is the time spent by the receiver to forward the payload to the application layer.
In this work we assume that $T_{rx} = 0$.
Also, as packets contain consensus updates, it is important to transmit the last generated update as soon as possible. 
Thus, we disable the transmission queue, i.e., we do not enqueue measurements that may have been produced while waiting for the PHY layer to become available for sending a new packet, and assume  $T_{queue} = 0$.

\subsubsection{Communication Protocol}

Each agent is producing a new status update every $T_{ca}$ seconds and it is assumed that all agents are synchronized.
The updates represent packets' payload and are sent to the transmission queue.
At the end of the transmission procedure, the agent stays in the receiving state until the next measurement arrives.
While in this state, it receives neighbors packets and delivers the received updates to the secondary control layer.

Clearly, if agents send the packet to the transmission queue at the same time instant, the probability of on-air collision is high.
Thus, we decorrelate the transmission instants  by adding a delay before sending the packet to the transmission queue, i.e., without modifying the MAC layer.
This delay can be random or deterministic; we decide for the latter approach and set its value $T_{AD}$ proportional to the agent's ID
\begin{equation}
\label{eq:ad}
T_{AD_{i}} = \sigma ( i - 1) .
\end{equation}
Thus, the total delay observed by the packet transmitted by the $i$-th agent becomes
\begin{equation}\label{addcf}
T_{total_i} = T_{AD_{i}} + T_{MAC}.
\end{equation}

\begin{figure}[!tb]
\centering
\subfloat[Channel without interference.]{\includegraphics[width=\columnwidth]{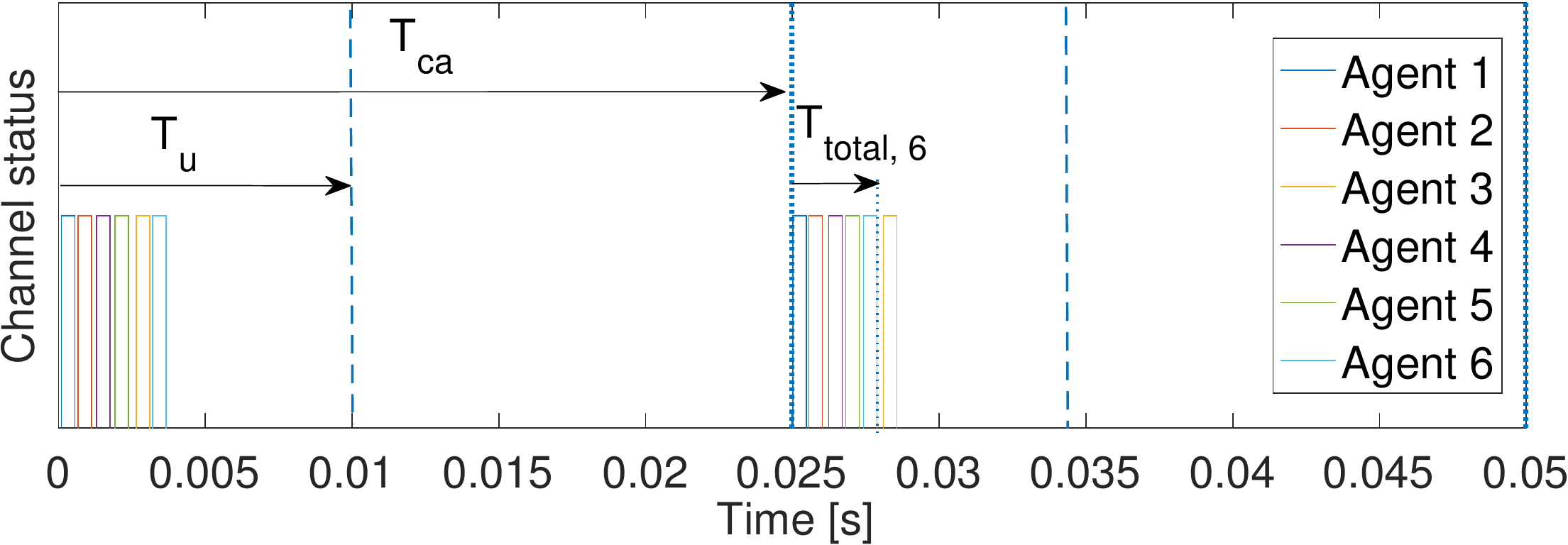}}
\hfil
\subfloat[Channel with unintentional interferer.]{\includegraphics[width=\columnwidth]{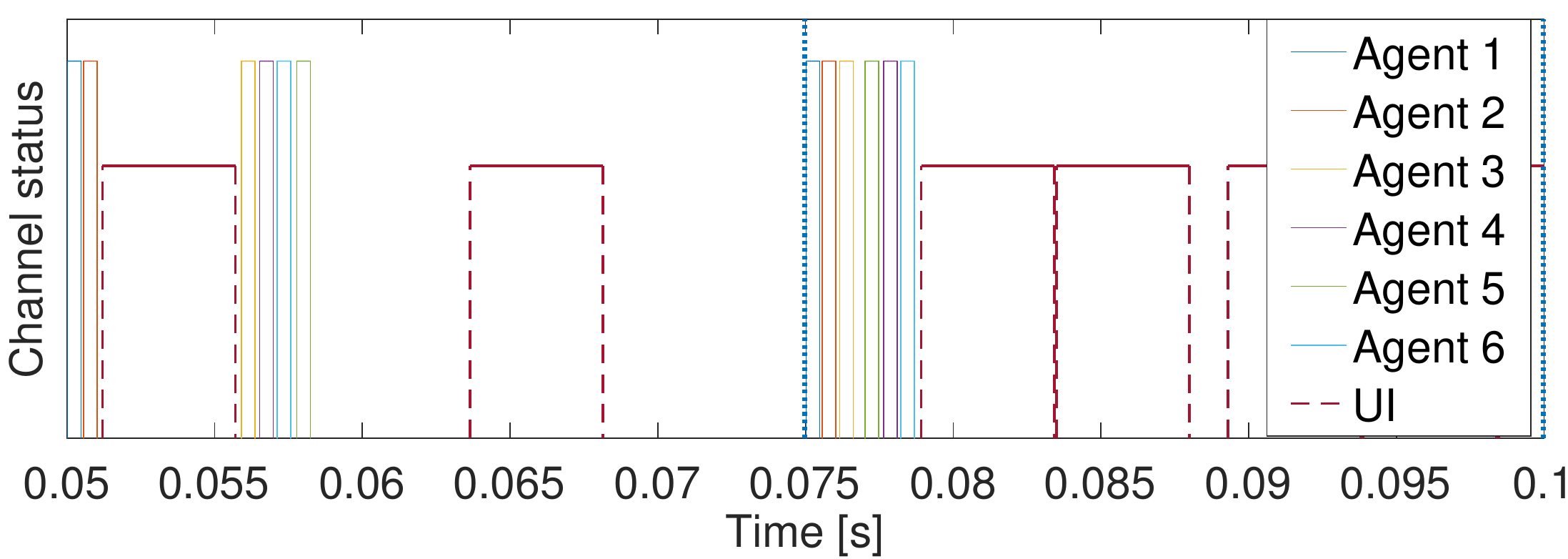}}

    \caption{The channel status seen by an external observer, $T_{ca} = 0.025 \, \text{s}$.}
    \label{fig:time}
\end{figure}

\subsubsection{Unintentional Interference Model}

While it is unrealistic to assume that the MG is communicating over a completely idle channel, it is also true that the interference pattern can be extremely variable among different environments.
In addition, the study of its impact is not the aim of this work.
Therefore, we use a simple model: the unintentional interferer (UI) is a 802.11 network where only one transmitter is producing downlink traffic.
The UI packets have constant lengths and arrive according to a Poisson process with parameter $\lambda$ to an infinite length transmission buffer.
The interferer follows the DCF rules in a fair way.
Fig.~\ref{fig:time} shows how the information exchange is seen by an external observer; for the way we have designed the protocol, it appears as a periodic packet burst. 


\subsubsection{Jammer Model}
\label{sec:jammer}

The attacker is using an off-the-shelf 802.11 device to transmit random packets, causing back-off freezing and collisions to the MG agents communication.
Note that if the jammer freezes \emph{all} agents and denies them communication, then the convergence is simply delayed and this can be detected rather easily.
Therefore, we assume that the output power of the jammer is limited, corrupting the communication of only a subset of agents.
In Section \ref{sec:attack} we show that such a simple jammer can make a great impact on the secondary MG control.

To reduce the number of jamming packets, we provide the jammer with a strategy that classifies it in the category of the reactive jammers.
Since the network is using broadcast packets, the MG is protected against control packet attacks, but the jammer can take advantage of the fact that the transmissions are periodic.
In fact, the MG communication appears to an external observer as a burst of $N$ packets every $T_{ca}$.
The period is continuously estimated as $\hat{T}_{ca}$ by the attacker simply sniffing the MG packets and accumulating their statistics.
The periodicity between two transmissions of an agent of each unit is averaged to minimize the error due to the CSMA/CA.
At the same time, the jammer prepares a new packet every $q \cdot \hat{T}_{ca}$ seconds.
The trigger for its transmission is the observation of a packet from any of the agents, as it signals the beginning of a packet burst, and $q \leq 1$ is a correction term added to have the trigger ready at the beginning of the burst.
The jamming packet is transmitted also if the channel is busy, i.e., the attacker is not respecting the DCF fairness.
This reactive jamming algorithm is simple and easy to implement, as it just needs the trace of received packets and produces as output a random packet sent on the wireless link.
To further reduce its detection probability, the jammer camouflages itself as the UI: this is achieved by using the same packet size and transmission scheme as the UI, as well as spoofing the UI's MAC address.

\begin{table}[t]
\centering
\caption{Control parameters.}
\label{table:parameters_control}
\begin{tabular}{|c|c|c|}
\hline
\multicolumn{3}{|c|}{\textbf{Primary control}}                                                                              \\ \hline
\textit{Virtual resistance} & \textit{Voltage PI loop} & \textit{Current PI loop} \\ \hline
$R_d=0.2 \, \Omega$          & $K_{pv}=4$, $K_{iv}=800$         & $K_{pc}=1$, $K_{ic}=97$         \\ \hline
\multicolumn{3}{|c|}{\textbf{Secondary control}}                                                                            \\ \hline
\textit{Consensus weight}   & \textit{Voltage PI loop} & \textit{Current PI loop} \\ \hline
$\epsilon = 0.025$          & $K_{psv}=0.02$, $K_{isv}=2$          & $K_{psc}=0.02$,  $K_{isc}=1$         \\ \hline
\end{tabular}
\end{table}

\section{Analysis}
\label{sec:analysis}

There are three aspects of the control that are related to the channel impairments: the \emph{sampling period}, the \emph{delay} and the \emph{reliability} \cite{lunze}.
The sampling period is $T_{ca}$ and corresponds to the rate at which a new status update is generated based on the local measurements.
The control delay, i.e., the time interval needed by an agent to collect all the neighbors samples, depends on the communication latency and bandwidth.
The values of $T_{u}$ and $T_{ca}$ are selected to guarantee an acceptably short control delay, without compromising the reliability.
In this work we set for simplicity $T_{u} = T_{ca}$; even if part of the agents could receive the packets from all neighbors before $T_{ca}$, they all compute \eqref{ca} at the same instant.

The channel unreliability can cause the absence of some of the neighbor's information at the input of the consensus block due to packet loss.
We refer to this event as \emph{feedback fault}.
It is necessary to define a strategy to adopt in case of missing feedback.
The topic is widely discussed in Networked Control System (NCS) literature \cite{drew}: in principle the strategy depends on the application, in particular how the system uses the output of the consensus algorithm.
The approach used in this work is the following. At $T_{u}$ the agent checks if all the neighbors information is received: if so, it can compute \eqref{ca} and save the result in a memory.
Otherwise, it discards all the received packets and outputs the value saved in the memory that corresponds to the last successful update period.
Another approach is to estimate the content of missing packets using the information received in the past, but it requires the study and evaluation of a proper method.\footnote{Our investigations showed that a simple strategy in which the content of a missing packet is replaced by the last successfully received value produces results that are worse than the ones presented in Section~\ref{sec:attack}. A more detailed study is left for future work.}


The feedback fault can happen only for a subset of the agents.
In this case at the decision instant only a part of agents will update their reference, while others continue keeping the previous one that was stored in the local memory.
Thus, we distinguish between two classes of feedback faults.
A \emph{coordinated fault} is detected by all the agents, and in practice can happen when the channel is busy for an entire $T_{ca}$, i.e.,  there is total absence of communication.
In this case, all the agents will continue to use the previous value as reference.
An \emph{uncoordinated fault} occurs when part of the agents received all the feedbacks and are able to compute (\ref{ca}), while the others are not able to do it because part of the feedback is missing.
The fault is caused by the fact that an agent cannot have the certainty that its neighbor received the packet because the communication loop is not closed, and the consequence is the loss of coordination among agents.
This is the kind of fault exploited by the jammer in our scenario.

\subsubsection*{Secondary Control Objective}

For the evaluation of the impact of the jamming attack, we need to clearly define the secondary control objectives, which are (i) the equal current sharing among DGs and (ii) the voltage restoration to $V_r$.
The imbalanced current sharing may result in overloading or even unintentional shut-down of some DGs, interrupting the system operation.
On the other hand, as the voltage deviation affects the performance of electric appliances, the MG voltage should be kept at nominal value.
Moreover, in case of an islanded MG, the over-voltage issue may cause inverse current flow to some of the DG units, inducing damages to the generation side.
Depending on the standardization, the tolerance on the deviation from $V_r$ may vary; in this work, we admit an absolute error of 5\% and 10\% on the steady state voltage, following the reference values used for MGs \cite{dragi}.
Moreover, in order to support the control objectives, the corresponding properties are requested from the consensus algorithm \cite{leblanc}:
\begin{itemize}
\item[P1] Termination: every DG decides for a state,
\item[P2] Validity: the voltage state estimated by each DG is $V_r$,
\item[P3] Agreement: every DG decides for the same state,
\end{itemize}
which we also investigate in the paper.

\section{The Impact of Jamming}
\label{sec:attack}

We investigate the impact of the jamming attack during the transient phase that follows the activation of the secondary control.
We base our study on the system used in \cite{meng}: the parameters used for the primary and secondary control are listed in Table \ref{table:parameters_control};  $T_{ca}$ is set to $25 \, \text{ms}$, guaranteeing the stability of the control; $V_r$ is fixed to $48~\text{V}$; there are $N=6$ DGs in the MG and the consensus topology is a full mesh, i.e., all agents are neighbors of each other.
The control is implemented in Simulink and uses PLECS software for the electrical model.



\begin{table}[]
\centering
\caption{Communication parameters.}
\label{table:parameters}
\begin{tabular}{|l|l|}
\hline
\textbf{Parameter}         		& \textbf{Value}    \\ \hline
Time slot duration $\sigma$     	& $20~\mu\text{s}$    \\ \hline
Propagation delay 			& $2~\mu\text{s}$     \\ \hline
Data rate         				& $1~\text{Mbps}$   \\ \hline
PHY duration      			& $96~\mu \text{s}$    \\ \hline
MAC header        			& $272~\text{bit}$ \\ \hline
DIFS              				& $32~\mu \text{s}$    \\ \hline
Contention window size $W$		& 32       \\ \hline
\end{tabular}
\end{table}

\begin{figure}[!t]
\centering
\includegraphics[width=0.9\columnwidth]{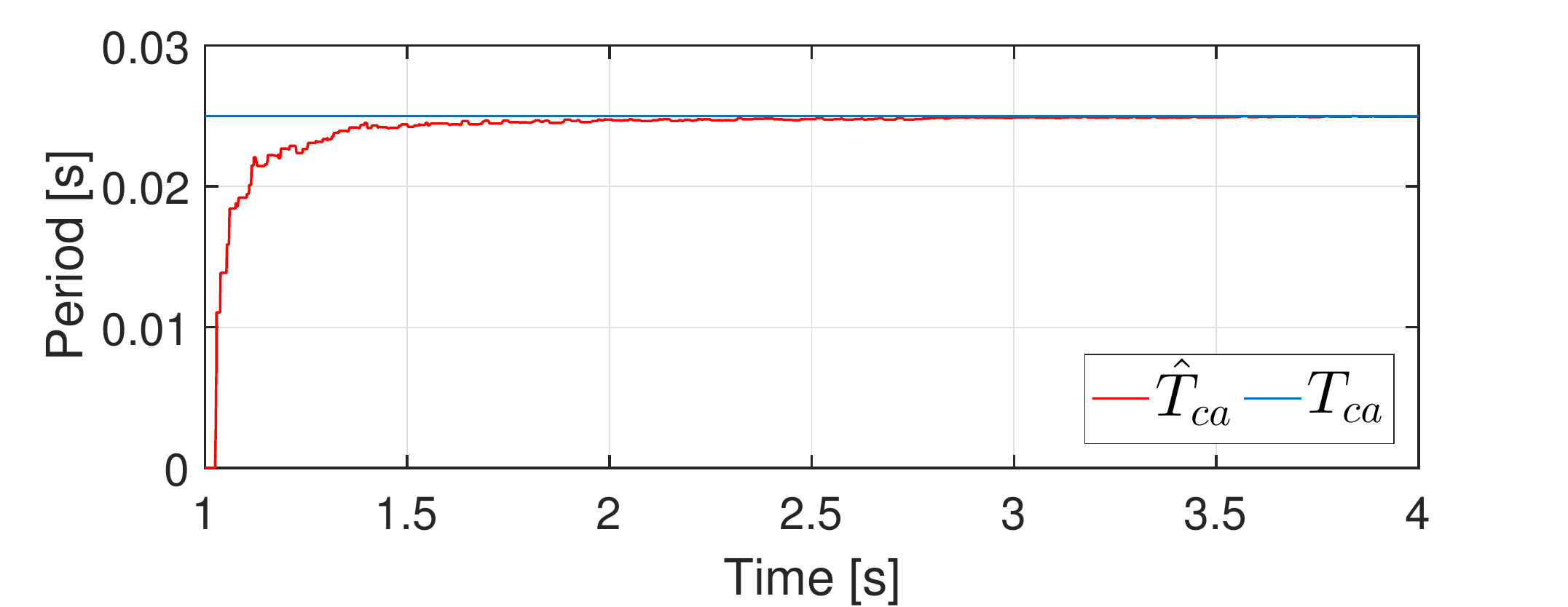}
\caption{Comparison between the sampling period estimated by the jammer and the actual one.}
\label{fig:tca}
\end{figure}

The communication system consists of 802.11g devices with parameters reported in Table \ref{table:parameters} and simulated in MATLAB.
The packet payload contains the consensus states of voltage and current, an incremental packet ID and encryption overhead, totaling 10 bytes.
The UI is also a 802.11g device that transmits with the same parameters as in Table~\ref{table:parameters}, with payload length of 512 bytes and packet arrival intensity of $\lambda = 100 \, \text{s}^{-1}$.
The jammer attacks agents 4 and 5, see Fig.~\ref{fig:control_loop}(a), waiting idle in their proximity until the first packet exchanged by the agents in the MG is captured.
Then it starts the attack, estimating in the same time the packet burst period $T_{ca}$; a realization of the estimation is plotted in Fig.~\ref{fig:tca}.
The value of $q$ (see Section~\ref{sec:jammer}) is set to 0.8.
If a packet is partially jammed, agents 4 and 5 are not able to decode it.
If it's corrupted by the collision with the UI or with another agent, no agent can decode it.
These are the two sources of packet loss.

\begin{figure}[!tb]
\centering
\subfloat[Consensus current states.]{\includegraphics[width=0.5\columnwidth]{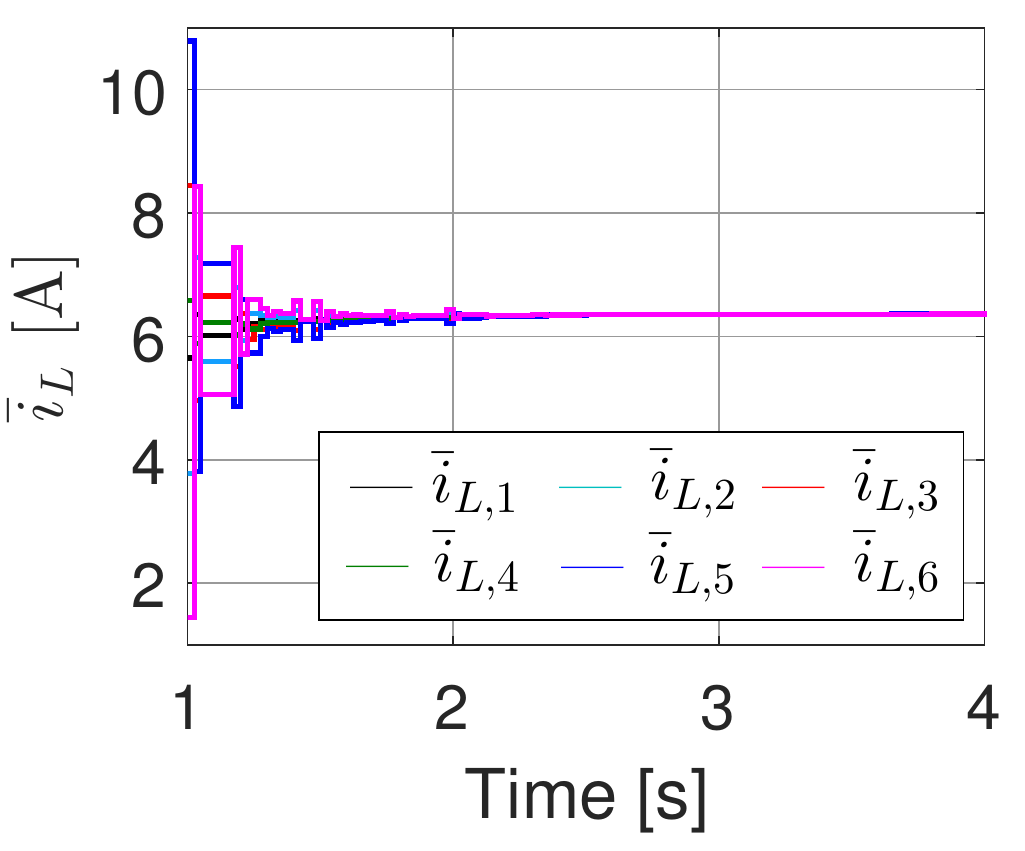}%
\label{fig:current_det}}
\hfil
\subfloat[Consensus voltage states.]{\includegraphics[width=0.5\columnwidth]{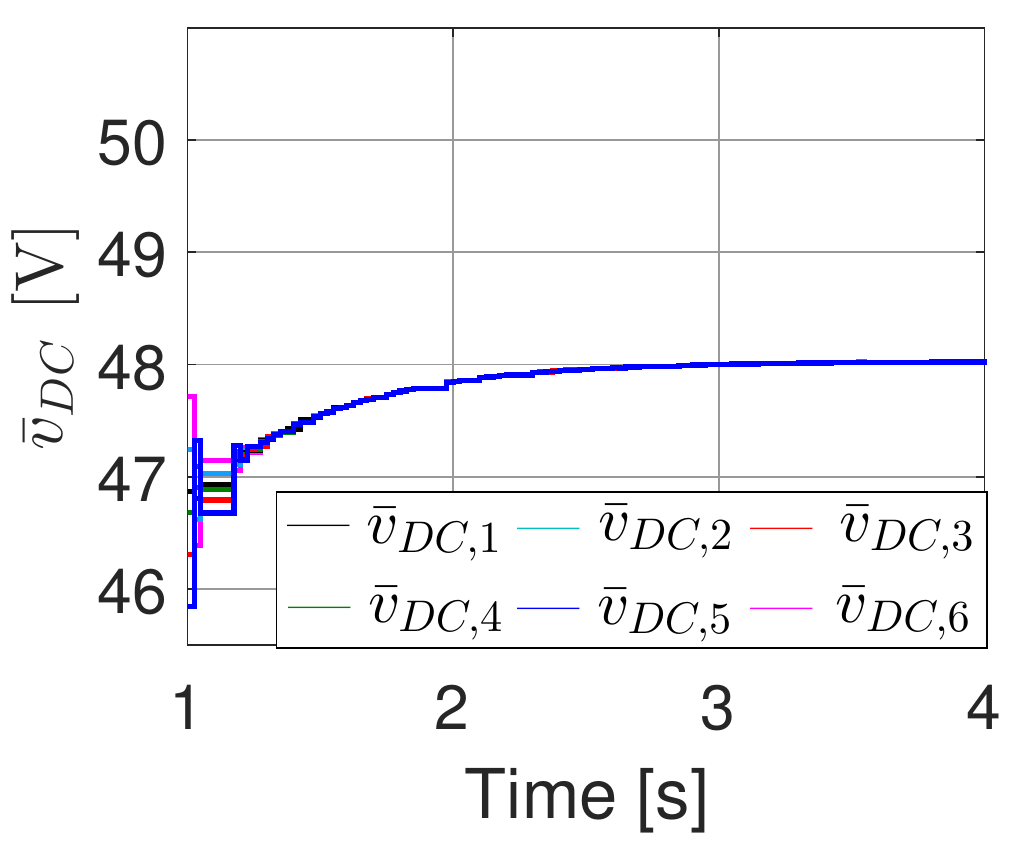}%
\label{fig:voltage_det}}

\subfloat[Consensus current states.]{\includegraphics[width=0.5\columnwidth]{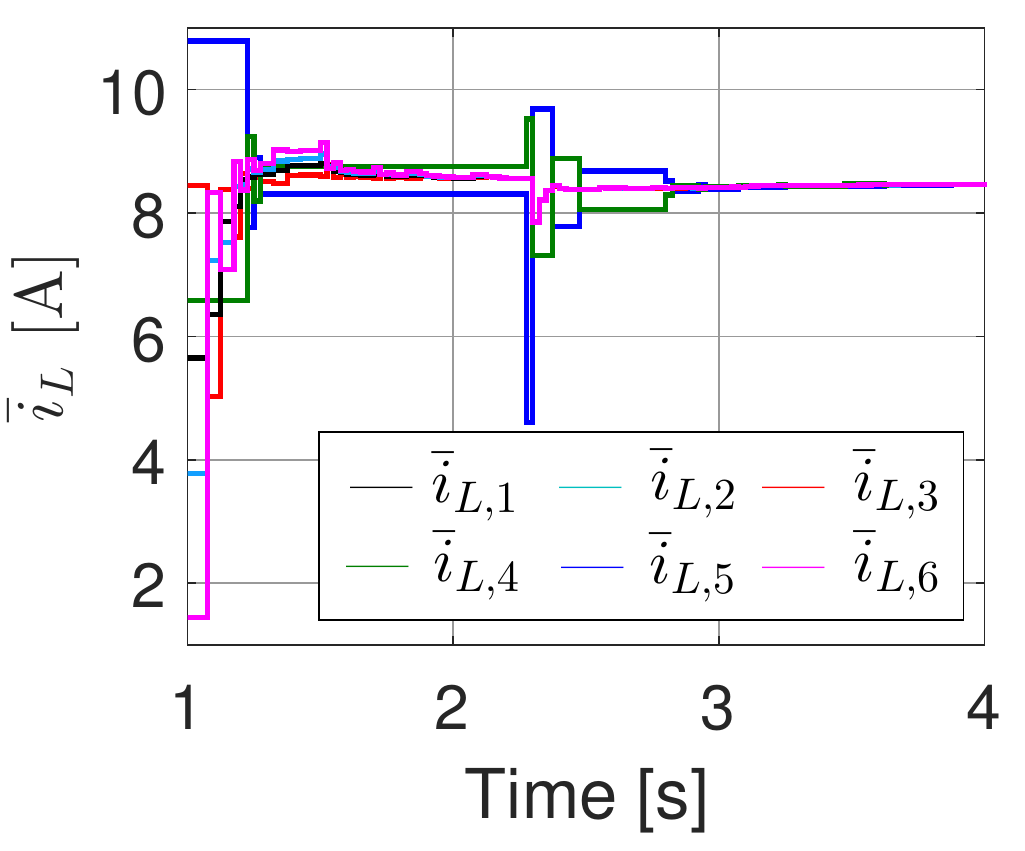}%
\label{fig:current_jam}}
\hfil
\subfloat[Consensus voltage states.]{\includegraphics[width=0.5\columnwidth]{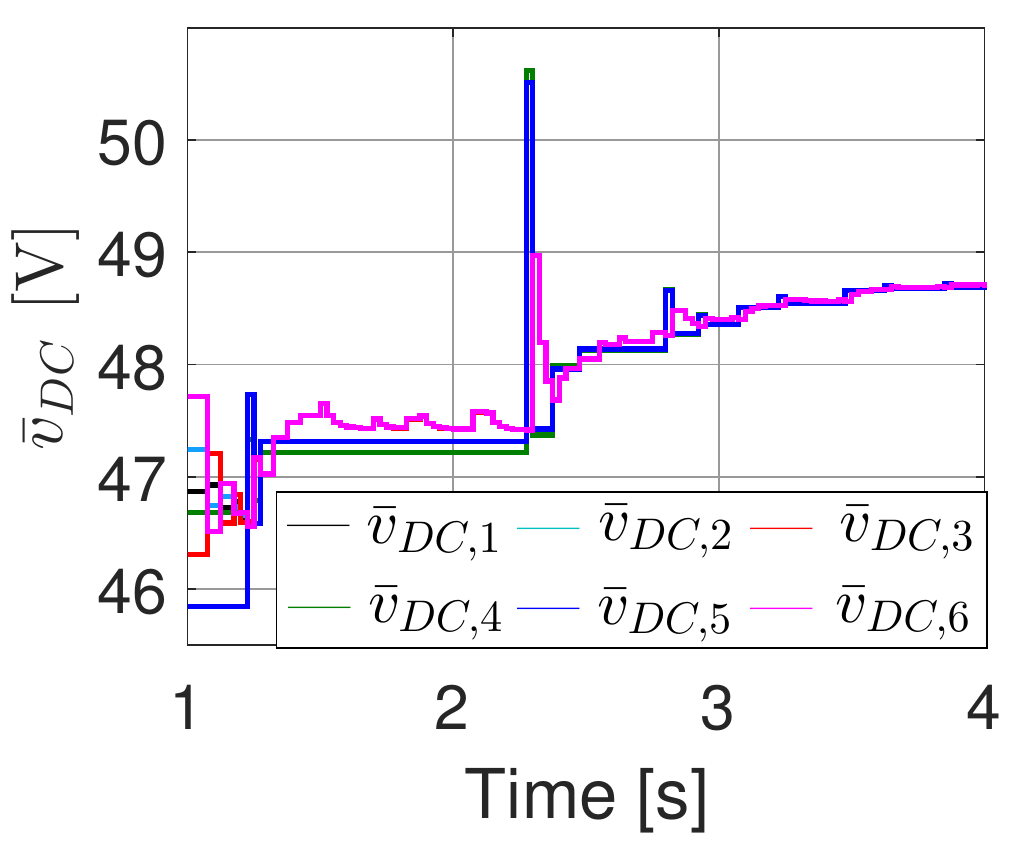}%
\label{fig:voltage_jam}}
\caption{An example of the evolution of the consensus states, (a)-(b) channel with UI only, (c)-(d) jammed channel.}
\label{consensus}
\end{figure}

\begin{figure}[!tb]
\centering
\subfloat[Channel with UI.]{\includegraphics[width=0.5\columnwidth]{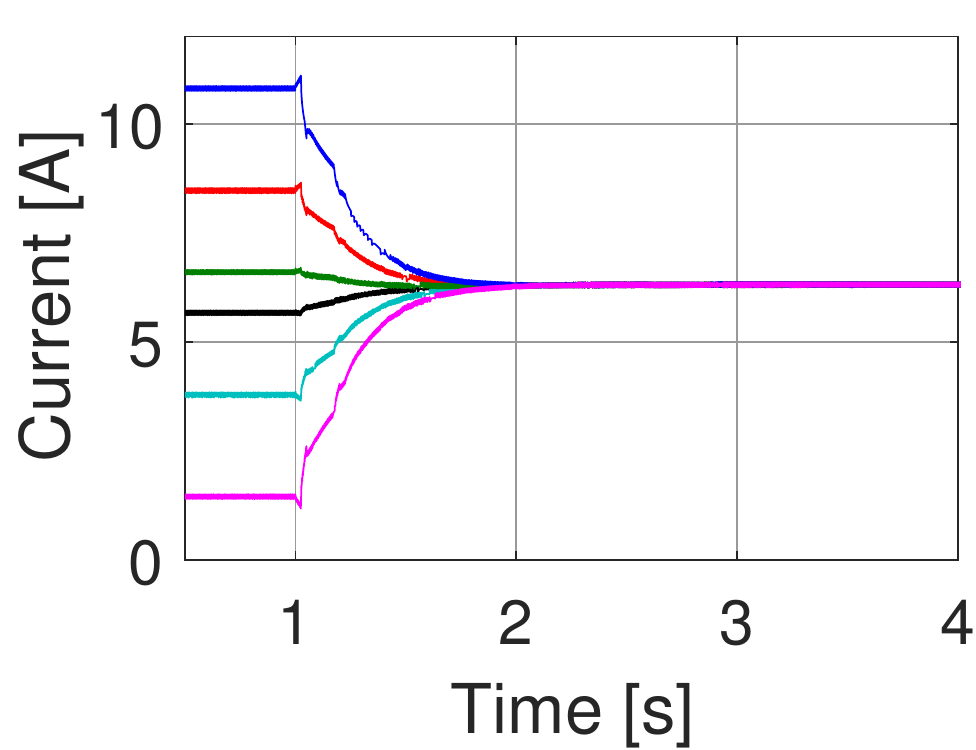}%
\label{fig:plecs_det}}
\hfil
\subfloat[Jammed channel.]{\includegraphics[width=0.5\columnwidth]{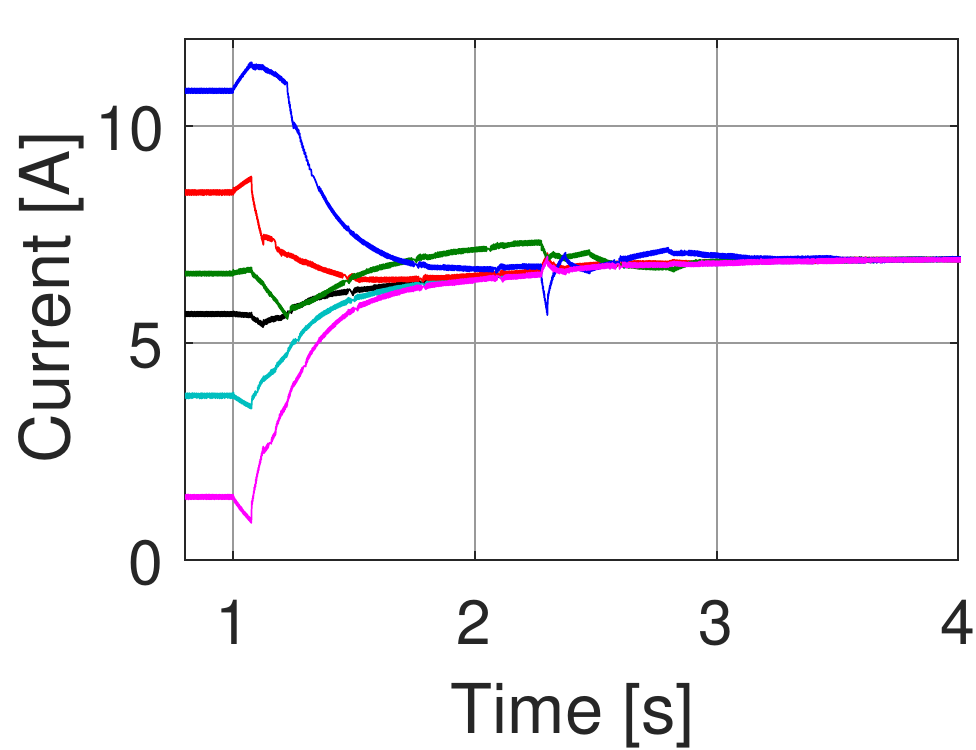}%
\label{fig:plecs_jam}}

\subfloat[Comparison of the voltage.]{\includegraphics[width=\columnwidth]{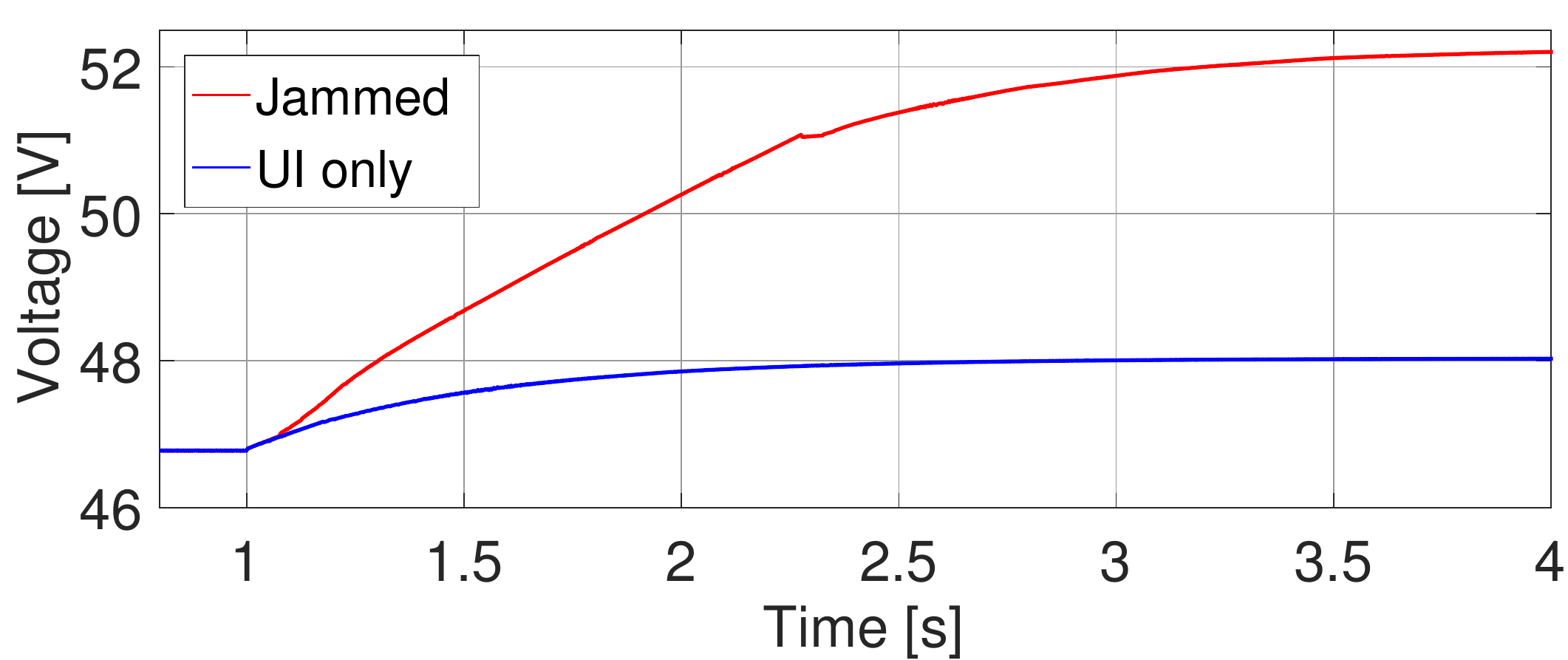}%
\label{fig:plecs_v}}

\caption{Simulation results obtained using PLECS.}
\label{plecs}
\end{figure}

Initially, only the primary control is active and the voltage is not the requested one.
At $t= 1\, \text{s}$ both the secondary control and the communications are activated and, after a transition period, a steady state is achieved.
We report in Fig.~\ref{consensus} the consensus states obtained from a simulation instance in a channel with the UI only and in a jammed channel.
Obviously, the system is able to cope with the UI, i.e., all properties P1-P3 are satisfied.
In case of jamming, the first observation is that the convergence speed is affected.
Regarding the steady state correctness, Fig.~\ref{plecs} shows the output current from each DG and the grid voltage for the same simulation instances.
Although the equal power sharing goal is achieved, grid voltage is restored to the wrong value of $52.25 \, \text{V}$, i.e., the error is $8.85$\%. 

\begin{figure}[!tb]
\centering
\subfloat[Mean voltage and voltage range.]{\includegraphics[width=0.5\columnwidth]{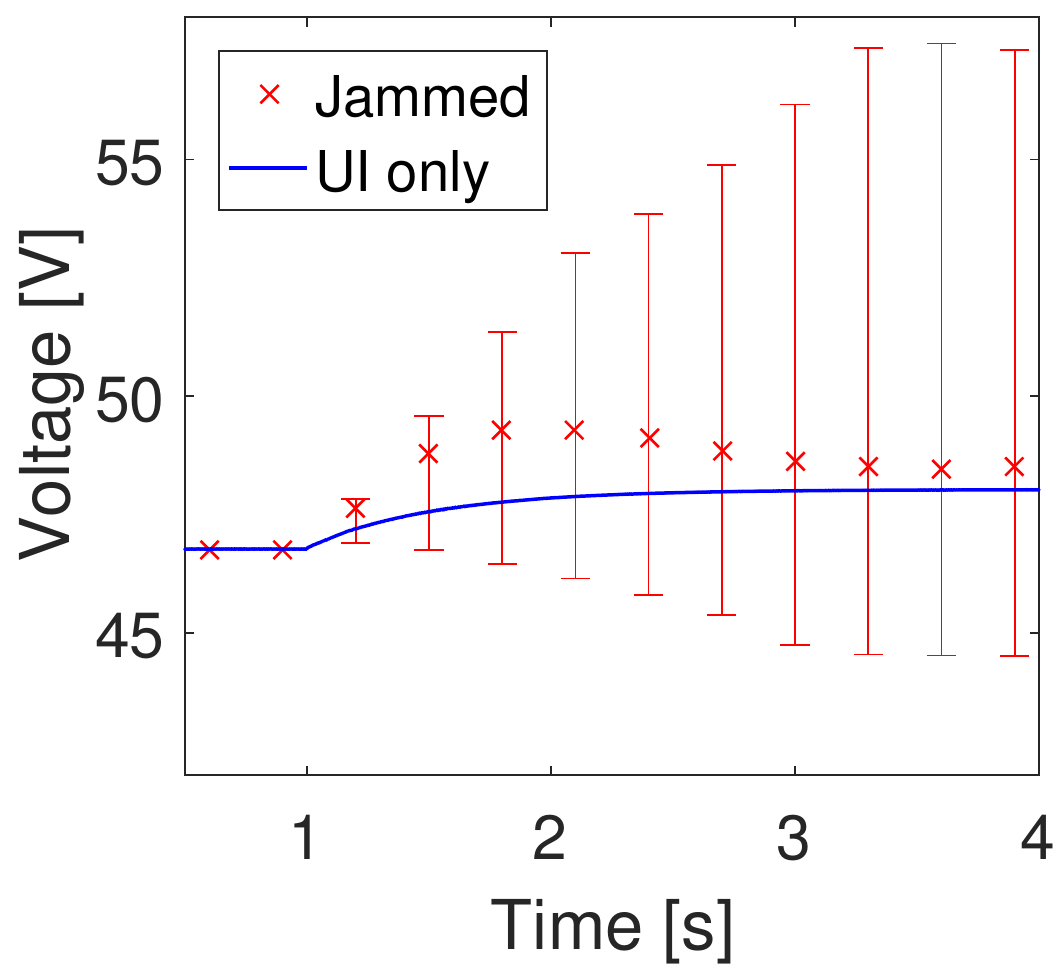}}
\hfil
\subfloat[Normalized histogram of the steady state voltage value.]{\includegraphics[width=0.5\columnwidth]{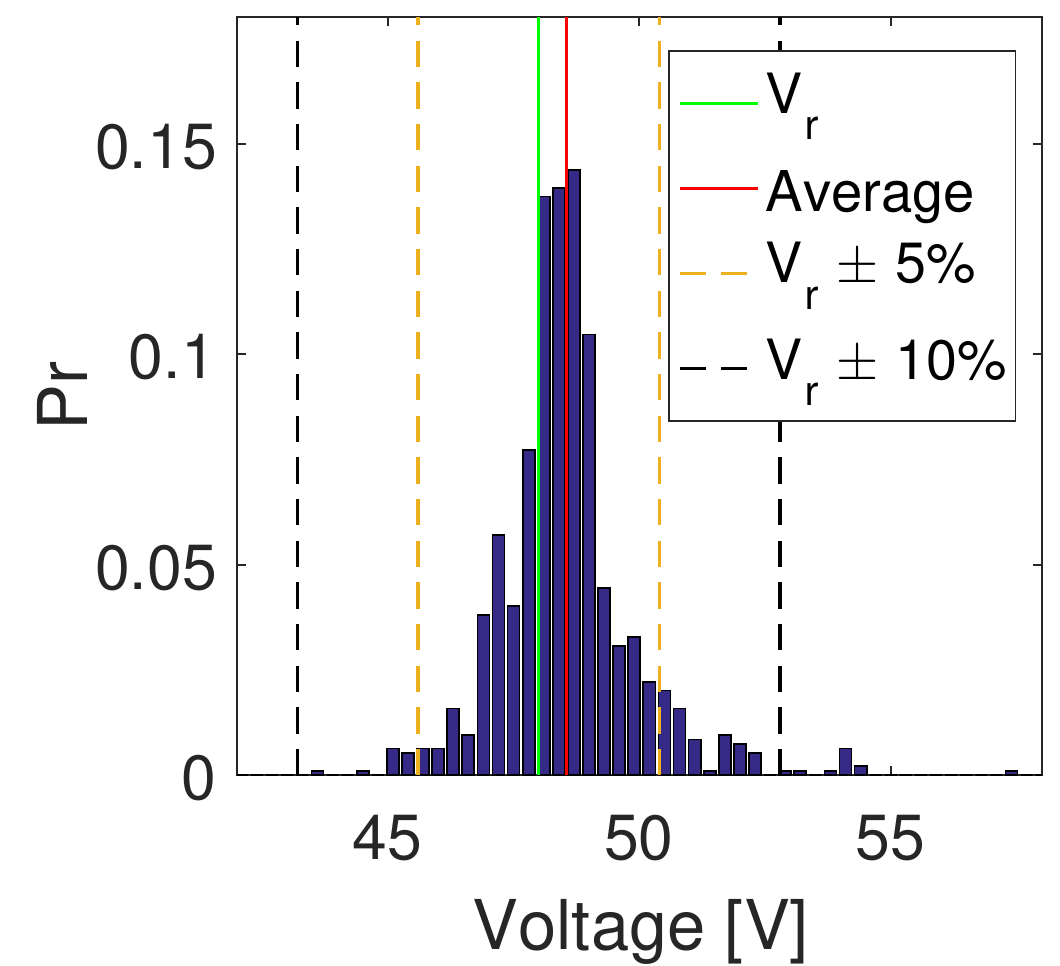}}
\caption{Simulation results.}
\label{res1}
\end{figure}

Fig.~\ref{res1} shows steady state voltage in the jammed scenario obtained over 1000 simulation runs.
Obviously, jamming significantly deviates the voltage, both during the transient phase, as well as when the steady state is reached.
In particular, the absolute error on the steady state voltage is greater than $5\%$ with probability $0.089$ and than $10$\% with probability 0.013.
These are significant values, especially as they are consequences of a simple jamming strategy.
Moreover, as the MG operation involves occasional load changes, when a new steady state has to be reached, there is a considerable potential to compromise MG stability with such a jammer.

\subsection{Possible Countermeasures}

\begin{figure}[!tb]
\centering
\subfloat[Mean voltage and voltage range.]{\includegraphics[width=0.49\columnwidth]{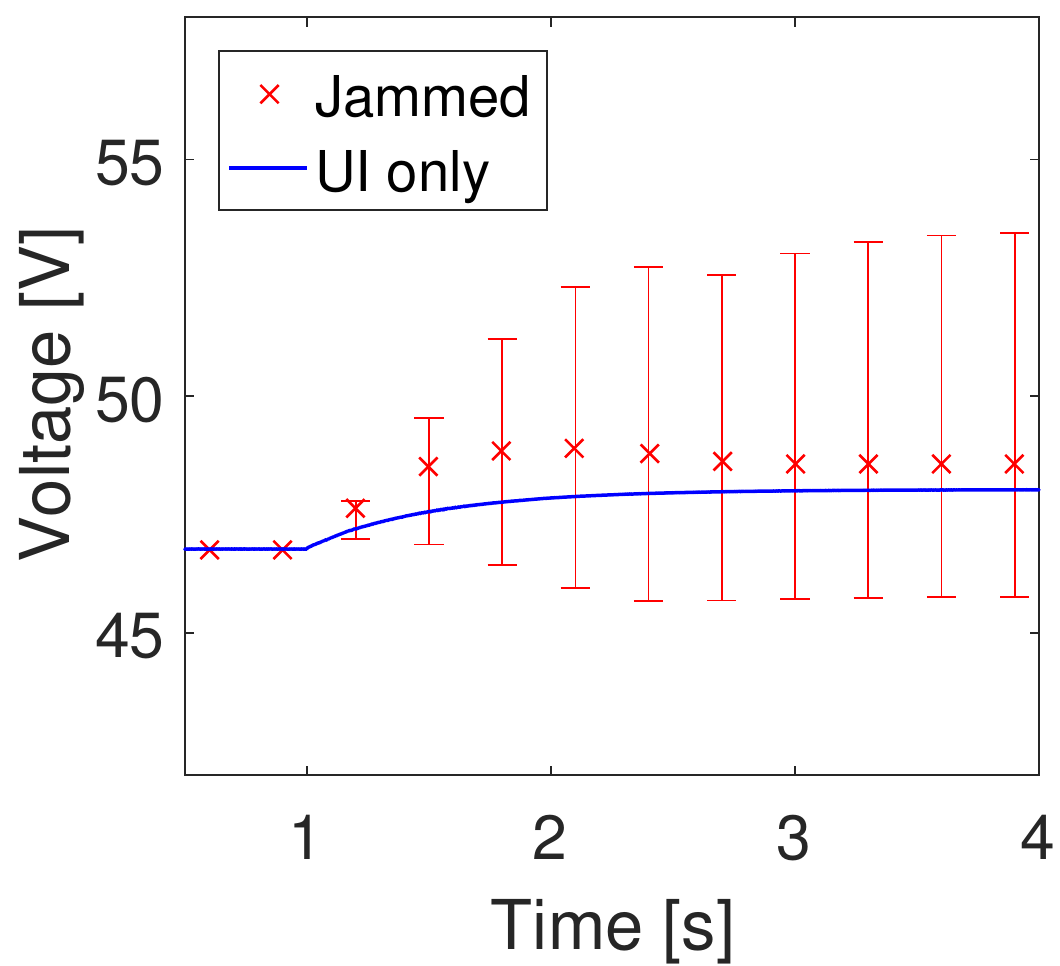}}
\hfil
\subfloat[Normalized histogram of the steady state voltage value.]{\includegraphics[width=0.49\columnwidth]{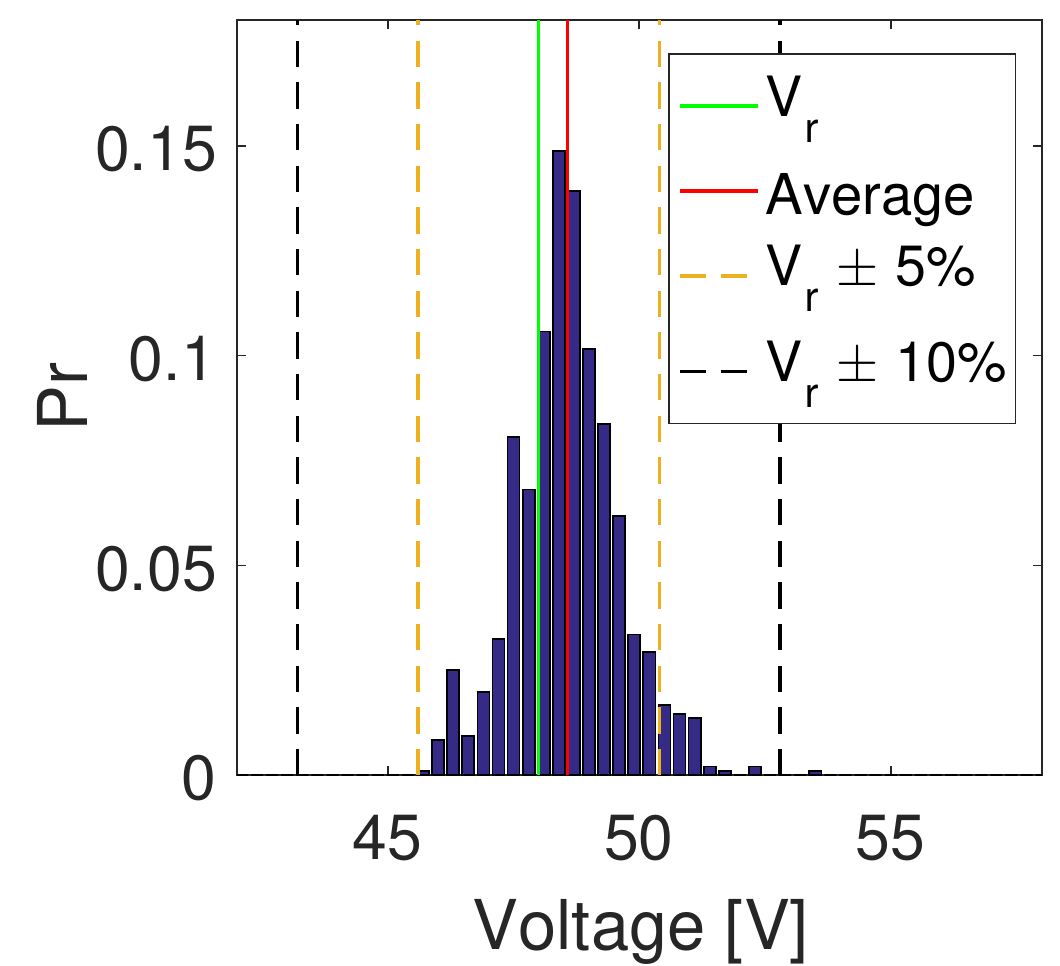}}
\caption{Simulation results.}
\label{res2}
\end{figure}

The disconnection of the attacked agents can be considered, but is clearly a drastic solution, as it decreases the energy offer: this motivates the need of a system adaptation.
The purposed solution is based on the fact that consensus interval is much longer than the on-air time of the exchanged packets; i.e., since $T_{ca} \gg N \, T_p$, we can spread the channel access by the agents during consensus interval to decrease the probability of jamming.
This is achieved modifying the $T_{AD}$, see \eqref{eq:ad}, and defining $T_e \leq T_{ca}$ as the part of $T_{ca}$ used for the AD.
We then divide $T_e$ in $N$ intervals and let the $i$-th agent transmitting in the interval corresponding to its ID.
The modified $T_{AD}$ is
\begin{equation}\label{mod_ad}
T_{AD_{i}} = ( i - 1) \, \frac{T_{e}}{N}.
\end{equation}
As the use of large $T_{e}$ increases the probability of status updates being received after the deadline $T_{ca}$ due to the CSMA/CA, its value should be carefully selected.
In Figure \ref{res2} we show the results obtained with $T_e = 12~\text{ms}$, a value that showed the most of the improvement in our investigations.
The corresponding results are presented in Fig.~\ref{res2}: the steady-state voltage value is greater than $48 \pm 5\%$ with probability $0.051$ and greater than $48 \pm 10\%$ with probability $0.001$.
Although this solution shows some benefits, they are modest and, in essence, inadequate.
A better protection may be obtained by more sophisticated methods, like communication anti-jamming techniques and fault tolerant consensus algorithms.

There are many countermeasures against jamming attacks in the communication literature, like frequency-hopping spreadspectrum (FHSS), direct-sequence spread-spectrum (DSSS), or even the use of directional antenna \cite{pelechrinis}, but they require coordination among agents.
Finally, the consensus literature offers interesting strategies that can be applied to the secondary control, e.g., by using broadcast gossip algorithms \cite{shafiee} or double-iteration algorithms \cite{garcia}.
As there is a trade-off between the communication complexity and the convergence speed \cite{olifati}, the selection of the optimal adjacency matrix in presence of jamming is an open problem.
In addition, a distributed mechanism is needed to check that the graph is strongly connected to guarantee the convergence.

\section{Conclusion}
\label{sec:conclusion}

In this paper we presented a potential weakness of the distributed MG control that has not received attention before.
Its impact is significant, as an attacker can strike the low-level functionality of the MG, i.e., the voltage restoration, using a simple jamming technique.
Indeed, if the secondary control communication is intended to rely on a wireless system, it should include adaptation and protection mechanisms.
Finally, we underlined how different components of the system can contribute to the mitigation of jamming attacks and motivated the future research about robust secondary control schemes.

\section*{Acknowledgment}
The work presented in this paper was partly supported by EU, under grant agreement no. 607774 "ADVANTAGE", and partly supported by the Danish Council for Independent Research grant no. DFF- 4005-00281 ``Evolving wireless cellular systems for smart grid communications''.




%

\end{document}